# A Bayesian algorithm for distributed network localization using distance and direction data

Hassan Naseri, *Student Member, IEEE*    Visa Koivunen, *Fellow, IEEE*

*Abstract*— A reliable, accurate, and affordable positioning service is highly required in wireless networks. In this paper, the novel Message Passing Hybrid Localization (MPHL) algorithm is proposed to solve the problem of cooperative distributed localization using distance and direction estimates. This hybrid approach combines two sensing modalities to reduce the uncertainty in localizing the network nodes. A statistical model is formulated for the problem, and approximate minimum mean square error (MMSE) estimates of the node locations are computed. The proposed MPHL is a distributed algorithm based on belief propagation (BP) and Markov chain Monte Carlo (MCMC) sampling. It improves the identifiability of the localization problem and reduces its sensitivity to the anchor node geometry, compared to distance-only or direction-only localization techniques. For example, the unknown location of a node can be found if it has only a single neighbor; and a whole network can be localized using only a single anchor node. Numerical results are presented showing that the average localization error is significantly reduced in almost every simulation scenario, about 50% in most cases, compared to the competing algorithms.

## I. Introduction

In recent years, the services that utilize the positions of network devices have become a key component of wireless systems [1]. We consider a wireless network comprised of anchor nodes with known locations, and target nodes with unknown locations to be estimated. Cooperative localization is a technique that employs communication among all the nodes to find the unknown locations [2]–[5]. That is, the measurements among target nodes are also utilized, in addition to the anchor-target measurements. Utilizing the extra information improves the identifiability of the localization problem [6]. Hence, cooperative localization can be applied to ad-hoc networks where not every node is connected to all the anchors. Propagation delays, i.e., time-of-arrivals (TOAs), and directions-of-arrival (DOAs) of radio signals amongst the network nodes can be estimated and employed for cooperative localization. High-resolution TOA and DOA estimation are facilitated by the increased bandwidth of radio signals along with a wide application of multi-antenna transceivers in most current wireless systems (e.g., 4G LTE and 802.11ac WiFi) and emerging technologies (e.g., 5G and evolution of WiFi) [7]–[9]. Hence, it is sensible to take advantage of the available angular domain information for wireless network localization in addition to delay data. Distance and direction data are two independent sources of information from different sensing modalities. Therefore, they can be combined to reduce the uncertainty in localization compared to distance-only or direction-only localization. Moreover, it helps to solve a localization problem with fewer anchor nodes and fewer connections among the nodes [4]. Practical applications include location based services (LBS) for 5G mobile networks, location-based routing and spectrum sharing for mobile devices, positioning of devices in WiFi networks, localization in sensor networks, and navigation of autonomous vehicles, robots, and first responders in emergency services. These applications are more important in indoor scenarios (including dense urban environments and covered paths), where conventional satellite-based and cell-based localization solutions may not be available or reliable. The main contributions of this paper are the following:

1) A new statistical model is developed for the problem of cooperative localization using hybrid distance and direction data. A likelihood function is derived to combine the joint statistics of distance and direction estimates using Gaussian and von Mises distributions. Both maximum likelihood (ML) and minimum mean square error (MMSE) estimators are formulated using the new data model.
2) The novel Message Passing Hybrid Localization (MPHL) algorithm is proposed to find an approximate solution for the formulated MMSE estimator. It is the first algorithm to approximate an optimal solution for the cooperative hybrid localization problem using a statistical estimation approach. It is a distributed algorithm based on belief propagation (BP) and Markov chain Monte Carlo (MCMC) sampling. Numerical results are provided to show that the Message Passing Hybrid Localization (MPHL) algorithm significantly reduces the localization error, typically 50% compared to the competing algorithms, in almost every scenario considered.

A theoretical study of cooperative localization using hybrid distance and direction data was reported in [4]. However, the results were based on noise-free observations, and the basic algorithm was only applicable to very specific network configurations. The Message Passing Hybrid Localization (MPHL) algorithm, proposed in this paper, employs noisy (erroneous) estimates of distance and direction quantities; and it is applicable to every rigid network configuration. The concept of rigidity will be discussed later, see [4] for more details. Distance-based cooperative localization has been studied in [5], [10]–[14]. In a fully connected network, the metric multi-dimensional scaling (MDS) [10] is an optimal algorithm for distance-only centralized localization [15, chap. 7]. Direction-only cooperative localization has been studied in [16], [17]. Cooperative localization methods in [2], [3] do not consider the joint statistics of distance and direction data, unlike the method developed in this paper.

A stochastic search algorithm for cooperative localization,

based on a Gaussian model for data and locations, was proposed in [18]. This model is not generally valid for hybrid localization, see Section II for more details. Although TOA, DOA, and received signal strength (RSS) measurements were suggested as input data in [18], only RSS-based results were presented. Recent advances in cooperative hybrid (distance and direction) localization were presented in [19], [20]. The CLORIS algorithm, proposed in [19], employed a heuristic cost function, which was not derived from a statistical data model. In [20], a Gaussian data model was proposed for hybrid data, The Gaussian distribution is not generally suitable for DOA estimates as it has an infinite support instead of a periodic support of $2\pi$ over angular domain [21]. Although an maximum likelihood estimator (MLE) was formulated in [20], the proposed algorithm, referred to as SDP1_Tomic, was not a direct approximation of the MLE. The statistical information, e.g., the error variances of the data, were discarded during the approximation. Moreover, the authors in [20] combined direction data with squared distance estimates, diverging more from the assumed model. The distributed versions of the CLORIS and SDP1_Tomic algorithms, proposed in [22], [23], were based on (block) coordinate descent. However, the exact order of the optimization (message schedule) were not specified. In the early stages of distributed cooperative localization, the node locations may have multi-modal distributions. Hence, a coordinate-descent method is only applicable to certain network configurations with sufficient anchor connectivity to avoid suffering from local minima problem. The proposed MPHL algorithm can be applied to any localizable network configuration, since it estimates and propagates probability distributions rather than node locations (single points).

The MPHL runs a sum-product message passing algorithm over a loopy factor graph model, a variant of the loopy belief propagation (LBP) [24], [25]. It is a distributed sequential algorithm, i.e., all the nodes update and propagate their beliefs (marginal probability distributions) in parallel. Distance-based cooperative localization algorithms stemming from LBP have been proposed in [12]–[14]. A multipath-aided hybrid localization method based on BP was proposed in [26]. They did not explicitly model distance and direction data, but the combined location error was modeled as normally-distributed. This model is not generally valid for hybrid localization, see Section II for details. The MPHL algorithm approximates BP messages using a set of samples (particles). It employs MCMC sampling to generate equally-weighted particles. An approximate MMSE estimate of a target location is obtained as the sample mean of it posterior distribution. Distance-based cooperative localization algorithms utilizing BP and importance sampling (weighted particles) were proposed [27]–[29]. In contrast to MCMC, importance sampling may suffer from the problem of sample degeneracy due to iterative re-weighting, and it requires the evaluation of exact posterior probability density functions (PDFs). The particle belief propagation (PBP) [30] and non-parametric belief propagation (NBP) [31] are general particle-base BP algorithms that employ MCMC sampling. The NBP applies kernel smoothing to all the messages (factors) to have well-defined products, which is not required by the PBP algorithm.

The proposed MPHL algorithm is a variant of the particle belief propagation (PBP) [30]. The main differences to the PBP include: (a) formulation of new factors (potential functions) for hybrid localization, (b) reduced complexity in communication, (c) improved numerical stability, and (d) employing a novel message scheduling mechanism designed for cooperative localization. These difference will be discussed later in more details.

The rest of this paper is organized as follows. Section II states the problem and data model. The Bayesian estimation framework, the factor graph model, and the proposed algorithm are described in Section III. Section IV considers the properties, requirements and extensions of the algorithm. Simulation results are presented in Section V. Finally, Section VI concludes the paper.

## II. Data Model

Assume a network comprised of $n$ nodes $\mathbf{x}_i \in \mathbb{R}^2$, $i = 1, \ldots, n$, from which $m$ are *target* nodes with unknown locations to be estimated, and $n - m$ are *anchor* nodes with known locations. The notation $\mathbf{x}_i$ is used to refer both to a network node and to its location. It is assumed that some pairwise distance estimates $r_{ij}$ and direction estimates $\alpha_{ij}$ among the nodes are available through TOA and DOA estimation techniques. The goal is to estimate the unknown locations of the target nodes using these observations. In 2D anchor-based localization, a single target node could be unambiguously localized using: (a) three TOAs to three different anchors, (b) two DOAs (with different values) to two different anchors, (c) a TOA and a DOA to a single anchor, (d) a TOA and a DOA to two different anchors that have a certain geometry, and (e) two TOAs and a DOA to three different anchors. The above conditions are sufficient using error-free observations with the exception of some degenerate cases, e.g., if three anchors lie on a single line. In cooperative hybrid localization with both (error-free) distance and direction estimates, all the locations can be found unambiguously using *a single anchor node* and *minimum connectivity* of the network, i.e., if the network graph is just singly connected. In a general cooperative hybrid localization scenario, i.e., different combinations of distance and direction estimates, the topology (connectivity) of the network determines if the problem is identifiable [4], [32]. That is, the network graph should satisfy certain rigidity requirements in order to unambiguously estimate all target locations, see [4] for more details. These rigidity requirements may be evaluated before starting the localization algorithm. The assumption in this paper is that, the localization problem has a unique solution with the given data.

The pairwise distances and azimuth angles between three nodes are shown in Fig. 1. A pairwise distance $d_{ij} = \|\mathbf{x}_i - \mathbf{x}_j\|_2$ is the Euclidean distance between the nodes $i, j$. The azimuth angle of node $j$ from node $i$, denoted by $\theta_{ij}$, is the angle of a vector from node $i$ to node $j$. An observed distance and direction at node $i$ are modeled as

$$\begin{aligned} r_{ij} &= d_{ij} + \epsilon_{ij}, \\ \alpha_{ij} &= \theta_{ij} + \gamma_{ij}, \end{aligned} \quad (1)$$

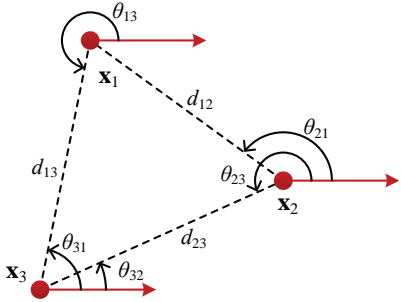

Fig. 1: Pairwise distances and directions

where $\epsilon_{ij}$ and $\gamma_{ij}$ are random error terms. These quantities are assumed to be obtained using high-resolution TOA and DOA estimation techniques. The estimation of these parameters are not in the scope of this paper; and *the estimated quantities are referred to as observations.* See [33]–[36] for examples of high resolution channel estimation techniques in wireless networks. It is assumed that the variances of distance and direction estimation errors are known or estimated reliably. These variances might be obtained from the measurements or approximated using performance bounds. It is also assumed that the distance and direction estimates are absolute quantities with respect to a general frame of reference. In order to get absolute distance estimates, the clock offsets of the nodes and other delays in the system should be compensated for, i.e., by time synchronization; see [5], [36] for more details. In order to have absolute direction estimates, i.e., with respect to a common coordinate axis, the orientations of all the nodes should be known.

Let us define three sets $\mathcal{H}, \mathcal{A}, \mathcal{X}$ that include all distance and direction estimates, and node locations respectively. The observations (obtained by a same node or by different nodes) are assumed to be conditionally independent, only depending on the locations of the two nodes involved. That is, the conditional PDFs of the observations, i.e., likelihoods, may be written as

$$f_{r_{ij}}(r_{ij} \mid \mathcal{H} \setminus r_{ij}, \mathcal{A}, \mathcal{X}) = f_{r_{ij}}(r_{ij} \mid \mathbf{x}_i, \mathbf{x}_j), \\ f_{\alpha_{ij}}(\alpha_{ij} \mid \mathcal{H}, \mathcal{A} \setminus \alpha_{ij}, \mathcal{X}) = f_{\alpha_{ij}}(\alpha_{ij} \mid \mathbf{x}_i, \mathbf{x}_j), \quad (2)$$

where \ denotes a set difference. Gaussian distribution is commonly used in the literature to model short-range distance estimation error [2], [37]. Hence, the conditional PDF of a distance observation with mean $d_{ij} = \|\mathbf{x}_i - \mathbf{x}_j\|_2$ and variance $\sigma_{ij}^2$ is given by

$$f_{r_{ij}}(r_{ij} \mid \mathbf{x}_i, \mathbf{x}_j) = \frac{1}{\sqrt{2\pi\sigma_{ij}^2}} \exp\left\{\frac{-1}{\sigma_{ij}^2}\left(r_{ij} - \|\mathbf{x}_i - \mathbf{x}_j\|_2\right)^2\right\}. \quad (3)$$

Directional data is usually modeled using von Mises distribution, which is a directional domain counterpart for Gaussian distribution [21], [38]. It provides an appropriate support for directional data, i.e., satisfies $f(\theta) = f(\theta + 2\pi k)$ for any integer $k$. The von Mises PDF fully defines the distribution of a DOA estimate using only two parameters, given by

$$f_{\alpha_{ij}}(\alpha_{ij} \mid \theta_{ij}) = \frac{1}{2\pi I_0(\kappa_{ij})} \exp\left\{\kappa_{ij} \cos(\alpha_{ij} - \theta_{ij})\right\}, \quad (4)$$

where $I_0(.)$ is the modified Bessel function of the first kind of order zero. The two parameters are the mean direction $\theta_{ij}$ (symmetry center) and the concentration parameter $\kappa_{ij} \in [0, \infty)$ (inverse scale), which are analogous to mean and variance in the normal distribution. The mean direction $\theta_{ij}$ corresponds to the true value of a direction parameter, i.e., the value of an error-free DOA estimate. If the concentration parameter is large, e.g., $\kappa_{ij} > 10$, it can be approximated as $\kappa_{ij} = 1/\zeta_{ij}^2$, where $\zeta_{ij}^2$ is the DOA estimation variance. Let us define two unit vectors $\mathbf{u}_{ij} = [\cos \alpha_{ij}, \sin \alpha_{ij}]^T$ as an observed direction vector, and

$$\mathbf{v}_{ij} = [\cos\theta_{ij}, \sin\theta_{ij}]^T = \frac{\mathbf{x}_j - \mathbf{x}_i}{\|\mathbf{x}_i - \mathbf{x}_j\|_2}, \quad (5)$$

as a true direction vector from node $i$ to $j$. The cosine of the angle between these two vectors is given by their inner product, i.e., $\cos(\alpha_{ij} - \theta_{ij}) = \mathbf{u}_{ij}^T \mathbf{v}_{ij}$. Hence, the PDF of a direction observation conditioned on the node locations may be written as

$$f_{\alpha_{ij}}(\alpha_{ij} \mid \mathbf{x}_i, \mathbf{x}_j) = \frac{1}{2\pi I_0(\kappa_{ij})} \exp\left\{\kappa_{ij} \mathbf{u}_{ij}^T \frac{(\mathbf{x}_j - \mathbf{x}_i)}{\|\mathbf{x}_i - \mathbf{x}_j\|_2}\right\}. \quad (6)$$

Using (3) and (6) and independence properties of the observations (2), the log-likelihood function for location parameters $\mathbf{X}$ is given by

$$\mathcal{L}(\mathcal{X}; \mathcal{H}, \mathcal{G}) = -\sum_{(i,j) \in \mathcal{U}} \frac{1}{\delta_{ij}^2}\left(h_{ij} - \|\mathbf{x}_i - \mathbf{x}_j\|_2\right)^2 \\ -\sum_{(i,j) \in \mathcal{V}} \kappa_{ij} \mathbf{u}_{ij}^T \frac{\mathbf{x}_j - \mathbf{x}_i}{\|\mathbf{x}_i - \mathbf{x}_j\|_2}, \quad (7)$$

where $\mathbf{x}_i$ are unknown parameters for $i = 1, \ldots, m$, and known anchor locations for $i = m+1, \ldots, n$. Index sets $\mathcal{U}$ and $\mathcal{V}$ contain index tuples $(i, j)$ for every observed distance and direction, respectively, e.g., if the distance between nodes $1, 2$ is observed at node 1 then $(1, 2) \in \mathcal{U}$. Assuming that unknown locations $\mathcal{X}$ are deterministic quantities, a ML estimate of the location parameters may be found by maximization the log-likelihood function (7). This is a highly non-convex problem that cannot be solved using conventional convex optimization techniques. Hence, a Bayesian formulation for the localization problem is proposed in the next section, that can be solved using approximate techniques. The approximation involves factorizing the joint estimation problem into several single-target localization problems, that can be solved using a sequential algorithm.

### III. BAYESIAN ESTIMATION

In the framework of Bayesian estimation, an unknown target location is treated as a random variable. In this section, an MMSE estimator is formulated for the hybrid localization problem; and the MPHL algorithm is proposed to solve it. It is an iterative Bayesian estimation algorithm that approximates the posterior distributions of target locations. The MPHL works by updating and propagating the marginal distributions of target locations over a factor graph model. The main reasons behind using this framework are as follows.

- An MMSE estimator can optimally combine the statistics of the distance and direction data. It can also handle multi-modal distributions when an ML or maximum a posteriori probability (MAP) estimator may suffer from local minima.
- Solving the MLE (7), or MMSE of hybrid localization, analytically or using convex optimization methods, is not a tractable problem. The factor graph approach breaks this problem into several small problems of single-node localization, that can be solved using a message passing algorithm.

## A. Minimum mean square error estimation

The MMSE estimate of a location variable $\mathbf{x}_i$ is the expected value of its marginal posterior distribution, given by

$$\hat{\mathbf{x}}_i = \mathbb{E}\left[\int f(\mathcal{H}, \mathcal{A} \mid \mathcal{X}) f^{(0)}(\mathcal{X}) \, \partial \mathcal{X} \setminus \mathbf{x}_i\right], \quad (8)$$

where $f(\mathcal{H}, \mathcal{A} \mid \mathcal{X})$ is the joint conditional PDF of the observations (likelihood function), $f^{(0)}(\mathcal{X})$ is a prior distribution on locations, and $\partial \mathcal{X} \setminus \mathbf{x}_i = \{d\mathbf{x}_1, \ldots, d\mathbf{x}_{i-1}, d\mathbf{x}_{i+1}, \ldots, d\mathbf{x}_n\}$ denotes integration with respect to all location variables except $\mathbf{x}_i$. The logarithm of the likelihood function $f(\mathcal{H}, \mathcal{A} \mid \mathcal{X})$ was given in (7).

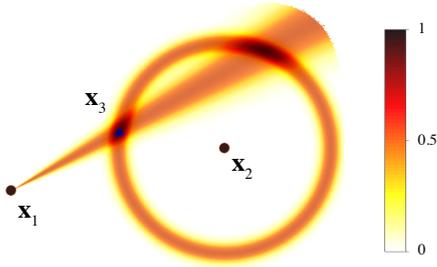

Fig. 2: An example of a posterior PDF for node 3 given the positions of nodes 1 and 2, and two observations. The direction $\alpha_{13}$ and the distance $r_{23}$ are observed. Combining distance and direction information reduces uncertainty about the location of node 3.

An example of a posterior distribution for the location of a single node $\mathbf{x}_3$ with uniform prior is illustrated in Fig. 2. There is one direction observation between $\mathbf{x}_1$ and $\mathbf{x}_3$, and one distance observation between $\mathbf{x}_2$ and $\mathbf{x}_3$. The locations of $\mathbf{x}_1$ and $\mathbf{x}_2$ are given. The locations of $\mathbf{x}_3$ may be found as the mean or maximum of this posterior distribution. The PDF in Fig. 2 has a complex shape with two modes. Multi-modal distributions arise in the context of cooperative localization due to: (a) insufficient number of anchor connections for a node, (b) large errors in TOA and DOA estimates, e.g., caused by multipath propagation. Many of the location parameters may have multi-modal distributions in the early stages of a distributed localization algorithm. By fusing more measurements and trough cooperation the uncertainty can be further reduced, and the PDF may become unimodal.

## B. Factor graph model

Factor graph is a graphical tool to represent the factorization of a joint probability distribution by exploiting the conditional independence properties of the variables. This factorization makes it possible to apply message passing algorithms to compute the marginals of an otherwise intractable joint PDF. Theoretical studies have proven the convergence and efficiency of such algorithms [39]. Moreover, the factor graph model for the cooperative localization problem directly maps to the communication network topology (nodes and links). Hence, it is a natural choice for this problem. A message passing algorithm can be easily distributed over a wireless network to run without a central control mechanism.

From (2), and assuming that the prior probabilities are all independent, the joint posterior probability may be factorized as a pairwise Markov random field (MRF). This factorization (up to a normalization constant) is given by

$$f(\mathcal{X} \mid \mathcal{H}, \mathcal{A}) \propto \prod_i \phi_i(\mathbf{x}_i) \prod_{(i,j) \in \mathcal{N}} \phi_{ij}(\mathbf{x}_i, \mathbf{x}_j). \quad (9)$$

The set $\mathcal{N} = \mathcal{U} \cup \mathcal{V}$ contains index tuples $(i, j)$ for every connected pair of nodes. A pairwise factor $\phi_{ij}(\mathbf{x}_i, \mathbf{x}_j)$ is (proportional to) the likelihood of the locations $\mathbf{x}_i, \mathbf{x}_j$ given the observations. A local factor $\phi_i(\mathbf{x}_i)$ is the evidence for node $i$. It is proportional to the prior probability of $\mathbf{x}_i$ multiplied by the probability of any local observation. If both distance and direction observations $r_{ij}, \alpha_{ij}$ between the nodes $i, j$ are available, then we have

$$\phi_{ij}(\mathbf{x}_i, \mathbf{x}_j) \propto f_{r_{ij}}(r_{ij} \mid \mathbf{x}_i, \mathbf{x}_j) f_{\alpha_{ij}}(\alpha_{ij} \mid \mathbf{x}_i, \mathbf{x}_j). \quad (10)$$

Since anchor locations are known, the observations between target nodes and anchor nodes can be incorporated into the definitions of local factors $\phi_i(\mathbf{x}_i)$.

The Bethe cluster graph is a simple graph model to represent a pairwise MRF [39, chap. 11]. For every factor or variable in (9) we create a vertex. Each variable is connected to all the factors sharing that variable. The result is a bipartite graph with the first layer for factors (clusters of variables) and the second layers for individual variables. This simple model can be constructed automatically. Fig. 3 shows an example of a factor graph corresponding to a network of four nodes with full connectivity. This is an undirected model assuming that the observations are reciprocal. That is, if $r_{ij}$ is observed then $r_{ji} = r_{ij}$ is also given. Similarly, if $\alpha_{ij}$ is observed then $\alpha_{ji} = -\alpha_{ij}$. The model and the rest of the results in this paper can be easily extended to the case of nonreciprocal observations.

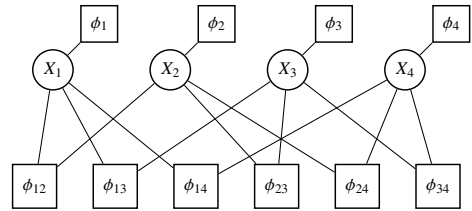

Fig. 3: Factor graph model for the joint probability distribution of the observations and variables. It is a bipartite graph and each pairwise factor is only connected to two variables.

The marginal posterior distributions of all variables in a factor graph can be computed by graph calibration using a sum-product message passing algorithm, also known as belief propagation [39, chap. 10-11]. In a Bethe cluster graph, due to the existence of loops, these marginals are computed using the loopy belief propagation (LBP) algorithm [24], [25]. The class of LBP algorithms perform approximate inference over a loopy graph by: (a) computing local beliefs (marginal posterior distributions) at each cluster/node using the messages received from its neighbors, (b) propagating the updated beliefs (messages) over the graph. This procedure is continued iteratively following a certain schedule until convergence, i.e., when the adjacent factors approximately agree on the distributions of the parameters. The convergence criteria of the MPHL algorithm will be discussed in the next section. An LBP algorithms may run in parallel or series over the clusters/nodes depending on the scheduling mechanism; and the parallel mode may require a centralized message synchronization. The MPHL algorithm employs an asynchronous scheduling mechanism, i.e., it runs in parallel over the nodes with no centralized control. The proposed scheduling mechanism will be described later. In the standard sum-product algorithm, i.e., BP, the messages are defined on a continuous state space. There are two types of messages corresponding to the graph model in Fig. 2: (a) a message from pairwise factor to a variable $\mu^{(t)}_{\mathbf{x}_i \to \phi_{ij}}(\mathbf{x}_i)$, and (b) a message from a variable to a pairwise factor $\mu^{(t)}_{\mathbf{x}_i \to \phi_{ij}}(\mathbf{x}_i)$, see [25] for details on the BP messages. At each iteration, the belief at node $i$, which is proportional to the marginal posterior probability of $\mathbf{x}_i$, may be obtained by multiplying all incoming messages at node $i$ with its local evidence.

### C. The proposed MPHL algorithm

The marginalization integrals of the joint posterior PDF in (9) cannot be computed using analytical methods. Hence, implementing an exact message passing algorithm on a continuous parameter space is not tractable. Several non-parametric message passing algorithms have been proposed in the literature to approximate LBP messages for continuous random variables [30], [31], [39, chap. 11]. The MPHL is mainly based on the particle belief propagation (PBP) algorithm [30], which obtains consistent estimates of the LBP messages. The differences to the PBP are as follows.

1) New factor functions are formulated for the MPHL, since the PBP has not been applied before to the problem of hybrid localization.
2) To reduce the communication cost: (a) each node broadcasts its current particle set instead of PBP messages; (b) these outgoing messages are sub-sampled.
3) All the factors in MPHL are computed in log-domain for numerical stability.
4) The convergence and results of LBP depend on the message order. While the PBP does not specify an order, the MPHL has a scheduling mechanism designed for cooperative localization. The fast convergence of this schedule is demonstrated by various numerical tests.

In the MPHL algorithm, the probability distributions of continuous variables are represented using finite sets of unweighted random samples. It avoids any biases associated with density estimation methods by using the particles to directly compute the messages. The MPHL algorithm is summarized in Table 1, and described in the following.

**Table 1:** Message passing hybrid localization (MPHL) algorithm

**Data:** Set of factors $\Phi$ and network graph $\mathcal{G}$
**Result:** Sampling distributions for each node location
// Create scheduling sets
1   $\mathcal{S}^{(t)} \leftarrow \text{createSchedule}(\mathcal{G}), \quad t = 1, \ldots, N_{iter}$
// Start message passing iterations
2   Anchor nodes transmit their locations $\mathcal{M}_{ij} = \{(\mathbf{x}_i, 1)\}$.
3   **for** $t = 1$ **to** $N_{iter}$ **do**
    // Receiving messages and sampling
4      All the nodes listen to incoming messages.
5      **for** *every node $i$ receiving messages* **do**
6         Construct a factor for each message received by node $i$, given by
        $\mu^{(t)}_{\phi_{ij} \to \mathbf{x}_i}(\mathbf{x}_i) \propto \sum_{(\hat{\mathbf{x}}_j, w_{ji}) \in \mathcal{M}^{(t-1)}_{ij}} \phi_{ij}(\mathbf{x}_i, \hat{\mathbf{x}}_j)/w_{ji}$,
7         Draw a new sample $\widehat{\mathcal{X}}^{(t)}_i$ of size $N$ from the current belief at node $i$, given by
        $b^{(t)}_i(\mathbf{x}_i) \propto \phi_i(\mathbf{x}_i) \prod_{j \in \mathcal{N}_i} \mu^{(t)}_{\phi_{ij} \to \mathbf{x}_i}(\mathbf{x}_i)$.
8      **end**
     // Transmitting messages
9      **for** *every node $i$ in the scheduling set $\mathcal{S}^{(t)}$* **do**
10        Select a random subset $\mathcal{Y}^{(t)}_i$ of size $M$ from $\widehat{\mathcal{X}}^{(t)}_i$.
11        **for** *every node $j$ in the neighborhood of $i$* **do**
12          Construct a message $\mathcal{M}^{(t)}_{ij}$ and transmit it to node $j$, given by $\mathcal{M}^{(t)}_{ij} = \left\{ (\hat{\mathbf{x}}_i, w_{ji}) \mid \hat{\mathbf{x}}_i \in \mathcal{Y}^{(t)}_i, w_{ji} = \mu^{(t)}_{\phi_{ij} \to \mathbf{x}_i}(\hat{\mathbf{x}}_i) \right\}$.
13        **end**
14      **end**
15   **end**

A message from a variable to a pairwise factor is an standard BP message, given by

$$\mu^{(t)}_{\mathbf{x}_i \to \phi_{ij}}(\mathbf{x}_i) = \phi_i(\mathbf{x}_i) \prod_{k \in \mathcal{N}_i \setminus j} \mu^{(t)}_{\phi_{ik} \to \mathbf{x}_i}(\mathbf{x}_i), \quad (11)$$

where $\mathcal{N}_i$ is the neighborhood of node $i$, i.e., the indices of its neighbor, and $\mu^{(t)}_{\phi_{ik} \to \mathbf{x}_i}(\mathbf{x}_i)$ is a message from a pairwise factor to a variable. This multiplication stage combines all the information about the location of node $i$ received from its neighbors except the message from node $j$. The belief at node $i$ at iteration $t$ is obtained by multiplying all incoming messages to node $i$ with its local evidence, as

$$b^{(t)}_i(\mathbf{x}_i) = \phi_i(\mathbf{x}_i) \prod_{j \in \mathcal{N}_i} \mu^{(t)}_{\phi_{ij} \to \mathbf{x}_i}(\mathbf{x}_i). \quad (12)$$

A local belief of a node is a finite-sample approximation of its marginal posterior PDF, up to a normalization constant. Using local beliefs, the equation (11) may be expressed alternatively as

$$\mu^{(t)}_{\mathbf{x}_i \to \phi_{ij}}(\mathbf{x}_i) \propto b^{(t)}_i(\mathbf{x}_i) / \mu^{(t)}_{\phi_{ij} \to \mathbf{x}_i}(\mathbf{x}_i). \quad (13)$$

Using this alternative form, a message from a pairwise factor to a variable is approximated by a sum over a set of samples

(particles) from its local belief $\hat{\mathcal{X}}_j^{(t)}$, given by

$$\mu_{\phi_{ij} \to \mathbf{x}_i}^{(t)}(\mathbf{x}_i) = \sum_{\hat{\mathbf{x}}_j \in \hat{\mathcal{X}}_j^{(t)}} \frac{\phi_{ij}(\mathbf{x}_i, \hat{\mathbf{x}}_j)}{\mu_{\phi_{ij} \to \mathbf{x}_j}^{(t-1)}(\hat{\mathbf{x}}_j)}. \quad (14)$$

where $t$ denotes the time instance, i.e., iteration number. This approximate marginalization computes the likelihood of a node location $\mathbf{x}_i$ given observations between nodes $i, j$ and the latest message from $\mathbf{x}_j$.

The particles are generated by direct sampling from the local beliefs of the nodes. The samples are drawn using the Metropolis-Hastings random walk MCMC algorithm [39, chap. 12] [40, chap. 6]. Since the samples are equally-weighted, the is no need for sample re-weighting. The Metropolis-Hastings algorithm can produce samples from a distribution by evaluating any function proportional to its PDF, i.e., node beliefs $b_i^{(t)}(\mathbf{x}_i)$. Hence, it does not require an extremely difficult computation of normalization terms. The algorithm works by drawing proposal samples from Gaussian distribution, which are then either accepted or rejected after evaluating the local beliefs $b_i^{(t)}(\mathbf{x}_i)$ analytically. The variance of the Gaussian proposal is tuned to achieve the desired acceptance rate for the samples. It has been shown that the ideal acceptance rate for a multi-dimensional target distribution is about 1/4 [41]. The lax requirements of the Metropolis-Hastings algorithm, makes it possible to implemented all the computation in log-domain for improved numerical stability. From (10), a logarithmic pairwise factor is given by

$$\phi_{ij}(\mathbf{x}_i, \mathbf{x}_j) = \frac{1}{\delta_{ij}^2} \left( h_{ij} - \|\mathbf{x}_i - \mathbf{x}_j\|_2 \right)^2 + \kappa_{ij} \mathbf{u}_{ij}^{\mathrm{T}} \frac{\mathbf{x}_j - \mathbf{x}_i}{\|\mathbf{x}_i - \mathbf{x}_j\|_2}. \quad (15)$$

The factor graph model in Fig. 3 directly maps to the structure of the underlying communication network. The clusters of parameter and measurement factors (vertices) are mapped to the network nodes, and the edges between them are mapped to the actual links of the wireless network. At every iteration, each network node collects messages form its neighbors to construct its local belief function $b_i^{(t)}(\mathbf{x}_i)$. Then it draws a new sample $\hat{\mathcal{X}}_i^{(t)}$ of size $N$ from its local belief. If the node is propagating this turn (depending on the schedule), then it selects a random subset of size $M < N$ from its current sample and broadcasts it to its neighbors along with the values of the incoming messages evaluated at sampling points. Thus, a transmit message $\mathcal{M}_{ij}^{(t)}$ is a set of tuples $\left( \hat{\mathbf{x}}_i, \mu_{\phi_{ij} \to \mathbf{x}_i}^{(t)}(\hat{\mathbf{x}}_i) \right)$. This iterative procedure is continued following a certain message passing schedule. The graphical model for the localization problem is a cyclic graph, see Fig. 3. Message passing in cyclic graphs, also known as loopy belief propagation, is an approximate inference method [25]. The final result and the convergence properties of the algorithm depend on the order of the messages. A dynamic scheduling mechanism is employed in the MPHL algorithm to guarantee the convergence. It starts by only anchor nodes transmitting their locations. The other nodes join the propagation schedule if they have already received a certain number of messages. The details of the scheduling and convergence properties of the algorithm will be discussed in next section. A final approximate MMSE estimate of a node location $\mathbf{x}_i$ is given by the the mean of $\hat{\mathcal{X}}_i^{(end)}$, i.e., the sample mean of it local belief at last iteration.

## IV. Properties of the algorithm

### A. Convergence

In the standard BP algorithm, a cluster graph is calibrated (the algorithm is converged) if every pair of adjacent clusters (measurements factors) agree on the distribution of their shared nodes [39, chap. 10-11]. Since an exact agreement can not be achieved in the LBP algorithm, a loopy graph may be considered calibrated if the factors approximately agree on the distributions of the parameters [42]. A particle-based BP algorithm can be stopped if the sampling distributions of the parameters are approximately constant, e.g., by checking the variation of the sample mean. That is, the localization algorithm can be stopped if the estimated locations of all the nodes remain approximately unchanged (withing a given error tolerance) in consecutive iterations. That is

$$\max \left\{ \|\hat{\mathbf{x}}_i^{(t)} - \hat{\mathbf{x}}_i^{(t-1)}\|_2 \mid i = 1, \cdots, m \right\} \leq \epsilon, \quad (16)$$

where $\epsilon$ is a given error tolerance for localization. However, evaluating such a criteria is not easy in the context of distributed localization. It requires communication between all the nodes before every iteration or a centralized control mechanism.

The stopping criterion for the proposed MPHL algorithm is not based on the graph calibration, but the maximum number of iterations determined by the message passing schedule. The convergence rate of the algorithm depends on the network size, the connectivity and geometry of the network, and the quality of the observations. The MPHL algorithm employs an automatic scheduling mechanism. The algorithm is stopped when the message passing schedule is finished. The proposed message passing schedule is described in the next subsection. The experiments show that, for uniquely localizable configurations considered in this paper, the MPHL algorithm converges very fast (e.g. in 10 iterations) and finds accurate estimates of the target locations. The convergence results are presented in the next section.

### B. Message scheduling

The MPHL algorithm runs iteratively until the message passing schedule is completed. There is no general message passing schedule to guarantee the convergence to exact marginals in a cyclic factor graph. However, a carefully designed schedule can ensure the convergence for an specific problem. In the problem of cooperative localization, an schedule can be designed in advance by analyzing the network connectivity, or it can be done dynamically by each node [43]–[45]. A dynamic scheduling mechanism employed by the MPHL, as described in the following.

1) First, only anchor nodes transmit their locations.
2) Later, target nodes also transmit their locations if they have received a certain number ($\gamma$) of messages in last iteration.
3) The schedule ends when last nodes in the schedule transmitted a certain number ($\nu$) of messages.

Setting $\gamma = 3$ guarantees that a target node propagates its location only if it can be uniquely determined. However, fewer connections, e.g., $\gamma = 1$ or $\gamma = 2$, might be sufficient for some nodes if both distance and direction data are available, as discussed in Section II. If a target node never receives $\gamma$ messages in a single iteration, it may still transmit if the total number of received messages equal $\gamma_{tot}$. This is necessary for cooperative localization in sparsely connected networks. In this paper, the value $\gamma_{tot} = 2\gamma$ is used to make sure that nodes with insufficient connectivity are adequately delayed in joining the schedule. The stopping number $\nu$ is proportional to the diameter ($\delta$) of the network graph [46], i.e., the greatest hop distance between any pair of nodes. In a cycle-free graph (tree), a distributed LBP algorithm converges after $\delta$ iterations [46]. In our experiments for loopy graphs, we set $\nu = 3\delta$. Hence, the length of the schedule (and the total number of iterations) depend on the network size and connectivity. An important advantage of the proposed scheduling mechanism is that, it does not require the message propagation and belief update stages over the network to be synchronized and centrally controlled. Hence, it allows for a fully autonomous and parallel execution of the localization algorithm over all the nodes. Although there is no theoretical guarantee for convergence of a loopy belief propagation algorithm, the numerical results show that the proposed scheduling mechanism provides good convergence properties in all the scenarios considered.

### C. Numerical stability

Sampling from the posterior distribution using a finite-precision arithmetics in a computer can lead to numerical stability issues. Evaluating a belief function for sampling, which includes products of many Gaussian and von-Mises densities, can easily overflow or underflow double precision arithmetic if the variances are very small. Underflow is a more severe problem, since it can prevent Markov chain from moving towards the modes of the distribution, unless initial state is in a region of high probability. To improve numerical stability the Metropolis-Hastings algorithm is implemented in logarithmic domain. Logarithmic transformation turns factor products to summations, helping to avoid underflows due to multiplication of very small probabilities.

### D. Computation complexity

The most computationally expensive part of the MPHL algorithm is sampling from posterior distributions, which runs at every node at every iteration. Hence, the overall computational complexity of the algorithm is determined by the complexity of the sampling mechanism. The following parameters impact the cost of sampling.
1) Sample size $N$: at each iteration every node draws a sample of size $N$ to construct its outgoing messages. Note that, constructing messages does not require extra computation since the corresponding weights are already calculated for sampling.
2) Neighborhood size $|\mathcal{N}|$: Drawing a single sample requires evaluating the product of incoming messages. The number of messages depends on the neighbors of each node. $|\mathcal{N}|$ can be the largest neighborhood size in the network.
3) Message size $M$: Each incoming message is a sum of $M$ functions (observation factors) corresponding to $M$ samples received from a neighbor.

Hence, drawing a single sample requires evaluating $O(|\mathcal{N}|M)$ likelihood functions. The complexity of sampling per iteration at a single node is $O(|\mathcal{N}|MN)$. Since the algorithm can be distributed over the network, each node can draws its samples locally. The total computation cost of the algorithm also depends on the number of message passing iterations required until convergence. The number of iterations in the schedule depends on the network configuration and connectivity. The schedule length does not directly grow by the network size, but is proportional to the network diameter (longest shortest path), i.e., the maximum hop distance between any two nodes in the network. The experiments show that, for the studied cases, the algorithm always converge in fewer than 10 iterations.

### E. Communication complexity

The complexity in communication, i.e., the total amount of data transfered among the nodes, depends on the
1) Message size $M$: at each iteration a node transmits $M$ samples to its neighbors.
2) Neighborhood size $|\mathcal{N}|$: messages from one node include different factor values (weights) for each of its neighbors. Hence, the message size depends on the number of its neighbors.
3) Neighborhood size $|\mathcal{N}|$: since the nodes in a single neighborhood transmit messages over a shared medium (in wireless networks), the amount of resources used for communication depend on the neighborhood size.

Hence, the total complexity in communication per iteration is $O(|\mathcal{N}|^2 M)$. Similar to the computation complexity, the total communication cost of the algorithm also depends on the number of message passing iterations required until convergence.

## V. Results

Numerical results for different stages of the algorithm are presented in this section. The results are produced using a simulated network of 10 nodes, from which 4 are anchor nodes. The network is partially connected, i.e., only some distance and direction observations are available. It is assumed that the observations are reciprocal. Fig. 4 shows an example of a simple network configuration. All distances are in meters but they can be easily scaled, as long as the measurement model is valid. In this example, there are 17 links between the nodes with both distance and direction observations. The standard deviation is 0.2 meters for distance observations and 10 degrees for direction observation. These error levels can be achieved using state of the art network technology, e.g., a multi-antenna WiFi system, and high resolution estimation algorithms [8], [9]. The results of the MPHL algorithm for nodes 5 and 9 are shown in Fig. 5 using only distance observations, and in Fig. 6 using both distance and direction observations (hybrid). These results are for 2, 7, and 10 iterations. The sample size is 1000, and the message size is 50. That is, in each iteration every node draws a sample of size 1000, and propagates 50 of them to its neighbors. It is evident from the network topology in Fig. 4 that none of the target nodes can be reliably localized using only anchor-target distance observations. However, as seen in Fig. 5, cooperative localization can provide accurate location estimates for all the nodes after few iterations. The sampling distributions of the locations change from spread-out multi-modal functions at iteration 1 (Fig. 5a, Fig. 5d) to sharp single-mode densities at iteration 10 (Fig. 5c, Fig. 5f). Fig. 6a shows that all the nodes in the considered scenario may be uniquely localized using only anchor-target distance and direction observations (combined), i.e., the beliefs are unimodal. However, as seen in Fig. 6b the uncertainty in location estimates can be significantly reduced through cooperation among the target nodes. The uncertainty regions significantly shrink after only 7 iterations. Comparing Fig. 5 and Fig. 6, shows that distance and direction data can be efficiently combined to reduce the uncertainty in localization.

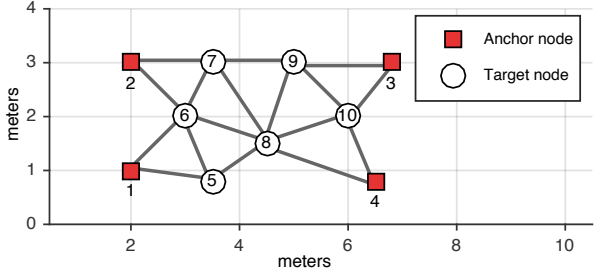

Fig. 4: Network model with four anchors and two target nodes, three distance observations and two direction observations.

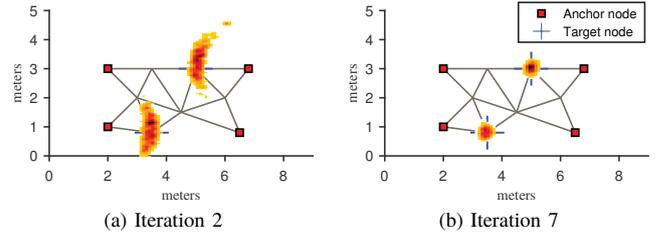

Fig. 6: Sampling distributions for target nodes 5 and 9 at iterations 2 and 7 using the MPHL algorithm with hybrid distance and direction data. The sampling distributions at iteration 2 are computed using anchor-target observations, as the target nodes are not transmitting any message yet. These initial distributions are spread out (high uncertainty) because each of these target nodes is directly connected to only one anchor. Each sampling distribution concentrates around a true target position after few iterations, i.e., the variance of the sample decreases. The sampling distributions at iteration 7 has single modes at approximately true target locations. The improvement in location estimates after few iterations, is obtained by cooperation between the target nodes.

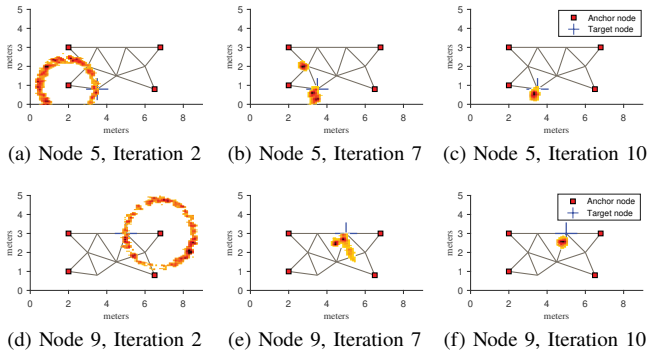

(a) Node 5, Iteration 2  (b) Node 5, Iteration 7  (c) Node 5, Iteration 10

(d) Node 9, Iteration 2  (e) Node 9, Iteration 7  (f) Node 9, Iteration 10

Fig. 5: Sampling distributions for target nodes 5 and 9 at iterations 2, 7, and 10 using the MPHL algorithm with only distance observations for the network configuration in Fig. 4. The sampling distributions at iteration 2 are computed using anchor-target observations, as the target nodes are not transmitting any message yet. The have circular shape because each of these target nodes (5,9) is directly connected to only one anchor. The sampling distributions gradually concentrate around true target positions after each iteration, i.e., the variance of the sample decreases. The sampling distributions at iteration 10 has single modes at approximately true target locations. The improvement in location estimates after few iterations, is obtained by cooperation between the target nodes.

The quality of localization results always depends on the network geometry, although the proposed approach makes it less dependent on the geometry compared to non-cooperative or non-hybrid methods such as satellite-based localization. For example the localization performance degrades if (a) three or more nodes (especially anchors) lie on a single line, (b) many nodes are outside the convex hull of the anchors, (c) the spatial distribution of the nodes is very nonuniform. The impact of network geometry on localization performance can be partially illustrated using geometric dilution of precision (GDOP) plots [47]. In order to control for the impact of network geometry, multiple random configurations are studied. In each configuration, 10 nodes are placed randomly in a $5\,\text{m} \times 10\,\text{m}$ area, with a constraint of $2\,\text{m}$ minimum separation between nodes, and four anchor nodes are randomly selected. Both distance and direction observations are available reciprocally for every connected pair of nodes. With 10 nodes, in a fully connected network there are 45 connection. However, the following results are produced for partially connected networks. A pair of nodes are connected if they are closer than $6\,\text{m}$. The resulting average connectivity is about 80% in the simulated networks, i.e., 36 connections in average. Multiple realizations of observation error is used to study each network configuration. Fig. 7 shows an example of a random network realization. The performance criteria for evaluating the results is the average localization error versus the standard deviations (STDs) of distance or direction estimates (observation error). The average localization error is the Euclidean distance be-tween the estimated and actual target locations averaged over all the target nodes in different network configurations with multiple realizations of observation error. The STDs of distance and direction estimates depend on different parameters of the measurement system and the environment, including the received signal-to-noise ratio (SNR), signal bandwidth, the size of a measurement packet (symbols), the number of antenna elements, the severity of multipath propagation, etc. The relationship between each term, e.g., SNR, and the observation error may be established using TOA and DOA estimation techniques or performance bounds [33], [35], [36].

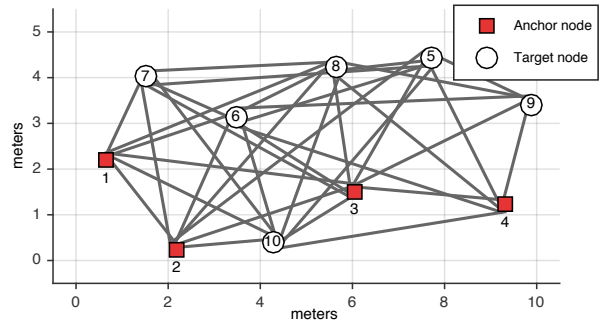

Fig. 7: Network model with four anchors and two target nodes, three distance observations and two direction observations.

Fig. 8a shows the average localization error in the network for the MPHL algorithm at different iterations using distance-only, direction-only and hybrid observations. The algorithm converges very fast and the localization error decreases monotonically. The algorithm converges in after only 5 iterations. In this scenario, the average localization error is reduced more than 20% using hybrid observations compared to the distance-only and direction-only cases. The standard deviation of sampling distribution (sum for all variables) is plotted in Fig. 8b. It shows that the uncertainty in location is decreasing over time and the hybrid localization provides more reliable results.

localization error of the proposed MPHL algorithm is less than 50% of the competing methods. Fig. 9b shows the average localization error of MPHL, SDP1_Tomic and CLORIS algorithms versus STD of direction estimates. The STD of distance estimates is 0.2 m in this simulation. The SDP1_Tomic performs better at higher levels of distance estimation error, while the CLORIS has its best performance at lower levels of angle estimation error. The results show that the MPHL algorithm outperforms both SDP1_Tomic and CLORIS at almost every error level. In these scenarios, the MPHL provides up to 30% of reduction in localization error compared to the competing methods. The better performance of SDP1_Tomic method compared to the MPHL at direction estimation STD of 10 m is due to the centralized processing in SDP1_Tomic compared to the distributed fashion of MPHL. Moreover, the results of the MPHL may be improved by increasing the number of particles. Fig. 10 shows the average localization error

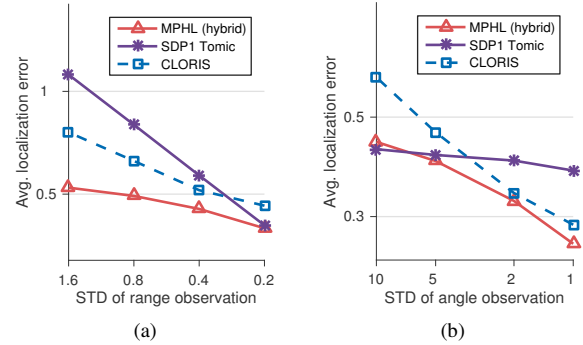

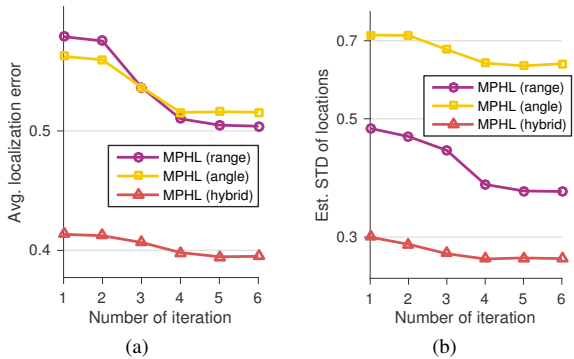

Fig. 8: Average localization error (a) and standard deviation of sampling distribution (b) versus number of iterations. STD of observations are 0.2 meters for distances and 5 degrees for directions.

Fig. 9: Average localization error (a) versus distance estimation error and (b) versus direction estimation error, in a partially-connected network. Number of iterations is 20. STD of angle observations is 5 degrees in (a). STD of distance estimation is 0.2 meters in (b). The proposed MPHL algorithm outperforms the competing hybrid localization methods at a wide range of observation error.

Fig. 9a shows the average localization error versus STD of distance estimates. The STD of direction estimation is 5 degrees in this simulation. The results of the proposed MPHL algorithm is compared with two recent hybrid localization algorithms, SDP1_Tomic [20] and CLORIS [22], [48]. These are the only two methods in the literature applied to hybrid data (distance and direction) for cooperative localization. Both the competing algorithms combine TOA and DOA observations using convex relaxation, i.e., the semi-definite programming (SDP) relaxation (SDP1_Tomic) and second-order cone programming (SOCP) relaxation (CLORIS). Message schedules for the distributed versions of SDP1_Tomic and CLORIS methods are not described in the corresponding articles. Hence, both the algorithms are implemented as centralized convex optimization methods using the CVX package [49]. The proposed MPHL algorithm outperform both SDP1_Tomic and CLORIS methods at a wide range of observation error. At higher levels of ranging error (STD of 0.5 meters), the average versus STD of distance estimates for the proposed MPHL algorithm, and metric MDS [10], SDP1_Tomic [20], and CLORIS [22], [48] algorithms. The STD of direction estimates is 5 degrees. These results are produced using the same randomly generated node configurations as above, but with full network connectivity. The metric MDS requires all pairwise distance observations to be available. It is shown that in this scenario, the MDS algorithm finds the optimal solution for distance-only localization [15, chap. 7]. The results show that the MPHL algorithm outperforms the competing hybrid localization methods and the metric MDS algorithm at a wide range of observation error. The average localization error of the MPHL is below the distance estimation error; and it is 20%–50% smaller compared to the competing algorithms. The improvement over the distance-only MDS comes from using hybrid observations. If distance estimates are sufficiently accurate, e.g., at STD of 0.2 meters, the MDS performs better than the SDP1_Tomic and CLORIS methods. However, the MPHL algorithm outperforms the MDS at every scenario by efficiently combining distance and direction data.

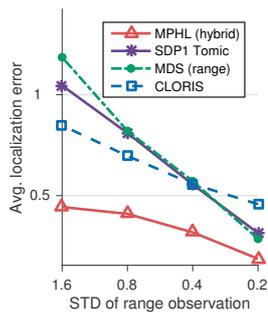

Fig. 10: Average localization error versus distance estimation error, in a fully-connected network. STD of direction estimates is 5 degrees. The proposed MPHL algorithm outperforms the competing hybrid localization methods and the metric MDS algorithm at a wide range of observation error.

## VI. Conclusions

In wireless networks, including WiFi and cellular networks, a reliable, affordable, and accurate positioning service is crucial. Due to the availability of both distance and direction information in modern wireless systems, it is essential to combine these two sensing modalities for wireless localization. The main contributions of this paper were: (a) the problem of cooperative network localization using hybrid distance and direction data was statistically modeled; and (b) the novel Message Passing Hybrid Localization (MPHL) algorithm was proposed to solve it. It is a fully distributed algorithm employing a novel scheduling mechanism. It efficiently combines the distance and direction data to find approximate MMSE estimates of target locations. Numerical results were provided to show the improvement in localization performance compared to existing distance-only and hybrid localization methods. For example, in the studied fully-connected networks of 10 nodes (4 anchors) with error variances of $1\,\mathrm{m}$ for distances and $5°$ for directions, the average localization error of the MPHL is about $45\,\mathrm{cm}$ compared to $70$–$95\,\mathrm{cm}$ for the competing algorithm. Possible directions for future work are to extend the algorithm for: (a) joint synchronization and localization, (b) joint estimation of locations and orientations of the nodes, and (c) to study and improve the scheduling mechanism for message passing.